\documentclass[11pt,preprintnumbers,aps,amssymb,nofootinbib,amsmath,superscriptaddress]
{revtex4}
\usepackage{epsfig,epsf}
\usepackage{bm} 
\usepackage{color} 
\newcommand{\beq}{\begin{equation}}
\newcommand{\beql}[1]{\begin{equation}\label{#1}}
\newcommand{\eeq}{\end{equation}}
\def\bal#1\gal{\begin{align}#1\end{align}}
\newcommand{\ball}[1]{\bal\label{#1}}
\newcommand{\eq}[1]{(\ref{#1})}
\newcommand{\fig}[1]{Fig.~\ref{#1}}

\newcounter{topiccounter}
\renewcommand{\b}[1]{{\bm #1}} 
\newcommand{\unit}[1]{\hat {{\bm #1}}} 

\newcommand{\e}{\varepsilon}

\begin{document}

\title{Quantum diffusion of electromagnetic fields of ultrarelativistic spin-half particles}

\author{Balthazar Peroutka}
\author{Kirill Tuchin}

\affiliation{Department of Physics and Astronomy, Iowa State University, Ames, Iowa, 50011, USA}

\date{\today}


\begin{abstract}

We compute electromagnetic fields created by a relativistic charged spin-half particle in empty space at distances comparable to the particle Compton wavelength. The particle is described as a wave packet evolving according to the Dirac equation. It produces the electromagnetic field that is essentially different from the Coulomb field due to the quantum diffusion effect.

\end{abstract}

\maketitle

\section{Introduction}\label{sec:intr}

It has been known for a while that very intense electromagnetic fields are created in ultra-relativistic hadronic and nuclear collisions \cite{Rafelski:1975rf,Greiner:1985ce,Ambjorn:1990jg,Olesen:2012zb,Ambjorn:1988tm}.  However, no convincing experimental evidence of their impact on the scattering dynamics has been observed. In recent years, a renewed interest to this subject was motivated by the relativistic heavy-ion collision experiments. The electromagnetic fields are intense enough to modify the properties of the nuclear matter produced in these collisions. In order to evaluate the impact of these fields on the nuclear matter, it is crucial to know their space-time structure. In \cite{Kharzeev:2007jp,Skokov:2009qp,Bzdak:2011yy,Voronyuk:2011jd,Ou:2011fm,Deng:2012pc,Bloczynski:2012en} production of the electromagnetic fields was studied using the hadron transport models, neglecting the nuclear medium electric and magnetic response and flow.  In \cite{Bzdak:2011yy,Deng:2012pc,Bloczynski:2012en} it was pointed out that the quantum nature of the nucleus wave function gives rise to strong field fluctuation, so that even in central collisions the r.m.s.\ does not vanish. In \cite{Tuchin:2010vs,Tuchin:2013apa,Tuchin:2013ie,Zakharov:2014dia} it is argued that due to the finite electric conductivity of nuclear matter, the lifetime of the electromagnetic field is significantly longer than in vacuum. Anomalous transport can also affect the field producing oscillations \cite{Tuchin:2014iua,Manuel:2015zpa,Li:2016tel} and even forcing the field into the topologically non-trivial configurations \cite{Tuchin:2016tks,Tuchin:2016qww,Hirono:2016jps,Hirono:2015rla,Xia:2016any,Qiu:2016hzd}. The electromagnetic field in the nuclear medium, unlike that in vacuum, strongly depends on the initial conditions \cite{Tuchin:2015oka}. The nuclear medium produced in relativistic heavy-ion collisions is well described by the relativistic hydrodynamics. Relativistic magneto-hydrodynamic calculations were done in \cite{Pu:2016ayh,Roy:2015kma,Inghirami:2016iru} in the ideal limit (infinite electrical conductivity).

In a recent publication \cite{Holliday:2016lbx} we argued that one can treat the sources of the electromagnetic field, i.e. the valence quarks, neither as point particles (classical limit) nor as plane waves, which have infinite spatial extent. This is because the interaction range, the quark wave function size and the dimensions of the produced nuclear matter have similar extent. As the first step towards understanding the quantum dynamics of the electromagnetic field sources, in \cite{Holliday:2016lbx} we modeled valence quarks as spinless Gaussian wave packets. Solving the Klein-Gordon equation  we computed the charge and current densities and the resulting electromagnetic fields in vacuum. In the present work we extend our approach to compute the spin contribution to the electromagnetic field. As in \cite{Holliday:2016lbx} we start with the initial Gaussian wave packet and evolve it in time according to the Dirac equation. At this point we completely neglect the medium effects as our goal is to study the effect of quantum diffusion of the quark wave function. This way our calculation is applicable to any hadronic and nuclear collisions.  

Before we proceed to the description of our calculation, it is worthwhile to set the notations, definitions  and normalizations.  The wave function of a spin-half particle can be expanded in a complete set of the momentum and helicity eigenstates:
\ball{a5} 
\Psi(\b r, t) =\frac{1}{\sqrt{2}} \sum_{\lambda}\int \frac{d^3k}{(2\pi)^{3/2}}e^{i\b k \cdot \b r}e^{-i\e_k t }\psi_{\b k}(0)u_{\b k\lambda}\,,
\gal
where $\e_k = \sqrt{m^2+k^2}$.  The four-component bispinor $u_{\b k\lambda}$ is the momentum and helicity eigenstate normalized as
\ball{a7}
u_{\b k\lambda}^\dagger u_{\b k\lambda'}= \delta_{\lambda\lambda'}\,.
\gal
$\psi_{\b k}(0)$ is the momentum wave function at $t=0$, normalized as
\ball{a9}
\int  |\psi_{\b k }(0)|^2 d^3k= 1\,.
\gal
With these conventions
\ball{a11}
 \int \Psi^\dagger(\b r, t) \Psi (\b r, t) d^3r = 1\,.
\gal
Solutions of the Dirac equation with given momentum $\b k$ and helicity $\lambda= \pm $ normalized by \eq{a7} are
\ball{a17}
u_{\b k +}=  \sqrt{\frac{\e_k+m}{2\e_k}}\left(\begin{array}{c} \chi_+ \\  \frac{\b \sigma\cdot \b k }{\e_k+m}\chi_+\end{array}\right)\,,
\quad 
u_{\b k -}= \sqrt{\frac{\e_k+m}{2\e_k}} \left(\begin{array}{c} \chi_- \\  \frac{\b \sigma\cdot \b k }{\e_k+m}\chi_-\end{array}\right)\,,
\gal
where the two-component spinors $\chi_\pm$ are  helicity eigenstates. 

\section{ Rest frame}\label{sec:b}

In the rest frame, although the particle momentum vanishes, the momentum of the Fourier components in \eq{a5} is finite, which is the reason for the wave function diffusion. Although the particle spin projection on any axis is conserved, only spin projection on the momentum direction is conserved for states with given momentum. This is why the helicity eigenstates are the correct choice of the spin basis.

Taking the direction of observation to be $z$-axis, i.e. $\b r = r\unit z$ and describing the momentum direction by the polar and azimuthal angles $\theta$ and $\phi$ we write the helicity eigenstates 
\ball{b5}
\chi_+= \left(\begin{array}{c}\cos\frac{\theta}{2} \\ \sin\frac{\theta}{2} e^{i\phi}\end{array}\right)\,,
\quad \chi_-= \left(\begin{array}{c}\sin\frac{\theta}{2} \\ -\cos\frac{\theta}{2} e^{i\phi}\end{array}\right)\,.
\gal
Using these in \eq{a17} yields
\ball{b7}
u_{\b k +}=  \sqrt{\frac{\e_k+m}{2\e_k}}\left(\begin{array}{c}  \cos\frac{\theta}{2}\\ \sin\frac{\theta}{2}e^{i\phi}\\ \frac{k}{\e_k+m}\cos\frac{\theta}{2}\\ \frac{k}{\e_k+m}\sin\frac{\theta}{2} e^{i\phi} \end{array}\right)\,, 
\quad 
u_{\b k -}=  \sqrt{\frac{\e_k+m}{2\e_k}}\left(\begin{array}{c}  \sin\frac{\theta}{2}\\ -\cos\frac{\theta}{2}e^{i\phi}\\ -\frac{k}{\e_k+m}\sin\frac{\theta}{2}\\ \frac{k}{\e_k+m}\cos\frac{\theta}{2} e^{i\phi} \end{array}\right)\,,
\gal
Plugging  \eq{b7} into \eq{a5} yields, after integration over the momentum directions (keeping in mind that $\b k\cdot \b r = kr \cos\theta$),  the wave function in the rest frame
\ball{b9}
\Psi(\b r, t)= \frac{1}{2 \sqrt{\pi}}\int_0^\infty dk k^2 e^{-i\e_k t }\psi_{\b k }(0) \sqrt{\frac{\e_k+m}{2\e_k}}
\left(\begin{array}{c} f(kr)  \\ 0 \\ \frac{ik}{\e_k+m} g(kr)\\ 0 \end{array}\right)
\gal
where 
\bal
f(z)&= \frac{1}{\sqrt{2}}\int_{-1}^1(\sqrt{1+x}+\sqrt{1-x})e^{izx}dx  \nonumber\\
&= 
\frac{\sqrt{\pi}}{z^{3/2}}\left\{ \sqrt{\frac{4z}{\pi}}\sin(z)-\cos(z) S\left( \sqrt{\frac{4z}{\pi}}\right) + \sin(z) C\left( \sqrt{\frac{4z}{\pi}}\right) \right\}
\,,\label{b11}\\
g(z)& =\frac{1}{i\sqrt{2}}\int_{-1}^1(\sqrt{1+x}-\sqrt{1-x})e^{izx}dx  \nonumber\\
&= 
\frac{\sqrt{\pi}}{z^{3/2}}\left\{ -\sqrt{\frac{4z}{\pi}}\cos(z)+\sin(z) S\left( \sqrt{\frac{4z}{\pi}}\right) + \cos(z) C\left( \sqrt{\frac{4z}{\pi}}\right) \right\}\,.\label{b12}
\gal
where $C$ and $S$ are the Fresnel integrals related to the error function:
\ball{b13}
C(z)+iS(z)= \frac{1+i}{2}\,\text{erf}\!\left[ \frac{\sqrt{\pi}}{2}(1-i)z\right]\,.
\gal
The corresponding charge and current densities are obtained using 
\ball{b14}
\rho= e\Psi^\dagger \Psi\,,\quad \b j = e\Psi^\dagger\b \alpha \Psi\,.
\gal
The resulting $\rho$ and $\b j$ are spherically symmetric. In particular $\b j$ is directed along  the $z$-axis, which coincides with the direction of $\b r$ in our coordinate system. 
\ball{b16}
\rho(\b r, t)&= \frac{e}{4 \pi}\int_0^\infty dk k^2 \psi_{\b k }(0) \sqrt{\frac{\e_k+m}{2\e_k}}
\int_0^\infty dl l^2 \psi_{\b l }(0) \sqrt{\frac{\e_l+m}{2\e_l}}\nonumber\\
  &\times \left\{ f(kr)f(lr)+ \frac{k}{\e_k+m}\frac{l}{\e_l+m} g(kr)g(lr)\right\} \cos[(\e_k-\e_l)t]\,.
\gal
\ball{b18}
\b j(\b r, t) &= \unit r \frac{e}{4 \pi}\int_0^\infty dk k^2 \psi_{\b k }(0) \sqrt{\frac{\e_k+m}{2\e_k}}
\int_0^\infty dl l^2 \psi_{\b l }(0) \sqrt{\frac{\e_l+m}{2\e_l}}\nonumber\\
  &\times \left\{ f(kr)\frac{l}{\e_l+m}g(lr)-  g(kr)\frac{k}{\e_k+m}f(lr)\right\} \sin[(\e_l-\e_k)t]\,.
  \gal

Given the charge and current densities, the electric field can be computed as
\ball{b20}
\b E(\b r, t)&= \int  \left\{ \frac{\rho(\b r',t') \b R}{R^3}+\frac{\b R}{R^2}\frac{\partial \rho(\b r',t') }{\partial t'}-\frac{1}{R}\frac{\partial {\b j}(\b r',t')}{\partial t'}\right\} d^3r'\,,
\gal
where  $t'=t - |\b r-\b r'|$ is the retarded time and $\b R= \b r-\b r'$. The only non-vanishing component of the electromagnetic field in the rest frame is the radial component of the electric field $E$:
\ball{b22} 
& \b E (\b r, t) = E(r, t)\unit r\,, & \b B(\b r, t) = 0\,.
\gal
We emphasize, that although the entire discussion of this section deals with a charged point particle at rest, its electromagnetic field $\b E (\b r, t)$ is different from the Coulomb field $e\b r/4\pi r^3$ due to the 
quantum evolution of the particle wave function. 

\section{Convective and spin currents}\label{sec:d}

It is instructive to separate the convective  and spin contributions (marked below by the subscripts $c$ and $s$ respectively).  Using the Gordon identity
\ball{d5}
\bar \Psi_2 \gamma^\mu \Psi_1= \frac{1}{2m}\left[ \bar \Psi_2 i\partial^\mu \Psi_1- (i\partial^\mu\bar \Psi_2)\Psi_1\right]-\frac{1}{2m}i\partial_\nu (\bar \Psi_2\sigma^{\mu\nu}\Psi_1)
\gal
we can write the charge and current densities \eq{b14} as
\ball{d7}
\rho = \rho_c+ \rho_s\,,\qquad \b j = \b j_c+\b j_s\,,
\gal
where 
\bal
\rho_c&= \frac{ie}{2m}(\bar \Psi \dot \Psi - \dot{\bar \Psi}\Psi)\,,\qquad  \b j_c= \frac{ie}{2m}\left[(\b \nabla \bar\Psi)\Psi- \bar \Psi(\b \nabla \Psi)\right]\,,
\label{d9}\\
\rho_s&= -\frac{e}{2m}\b \nabla \cdot (\Psi^\dagger\b \Sigma \Psi)\,,\qquad
\b j_s= \frac{e}{2m}\partial_t(\Psi^\dagger \b \Sigma \Psi)-\frac{ie}{2m}\b\nabla \times (\Psi^\dagger\gamma_0\b \Sigma \Psi)\,.
\label{d10}
\gal
Explicit expressions for the convective charge and current densities read
\ball{d12}
\rho_c(\b r, t)&= \frac{e}{8 \pi m}\int_0^\infty dk k^2 \psi_{\b k }(0) \sqrt{\frac{\e_k+m}{2\e_k}}
\int_0^\infty dl l^2 \psi_{\b l }(0) \sqrt{\frac{\e_l+m}{2\e_l}}\nonumber\\
  &\times (\e_k+\e_l)\left\{ f(kr)f(lr)+ \frac{k}{\e_k+m}\frac{l}{\e_l+m} g(kr)g(lr)\right\} \cos[(\e_k-\e_l)t]\,.
\gal
\ball{d13}
\b j_c(\b r, t) &= \unit r \frac{e}{8 \pi m}\int_0^\infty dk k^2 \psi_{\b k }(0) \sqrt{\frac{\e_k+m}{2\e_k}}
\int_0^\infty dl l^2 \psi_{\b l }(0) \sqrt{\frac{\e_l+m}{2\e_l}}\sin[(\e_l-\e_k)t]\nonumber\\
  &\times \left\{l f(kr)f'(lr)-kf'(kr)f(lr)+\frac{l}{\e_l+m}\frac{k}{\e_k+m}\left[lg(kr)g'(lr)-kg'(kr)g(lr)\right]
\right\} \,.
 \gal
Here the integrals 
\bal
f'(z)&= -\frac{1}{\sqrt{2}}\int_{-1}^1(\sqrt{1+x}+\sqrt{1-x})x\sin(zx) dx\,,\label{d15}\\
g'(z)&= \frac{1}{\sqrt{2}}\int_{-1}^1(\sqrt{1+x}-\sqrt{1-x})x\cos(zx) dx\label{d16}
\gal
can be expressed in terms of the Fresnel integrals \eq{b13}. The corresponding field is computed using \eq{b20}. 

\section{Lab frame}\label{sec:c}

In the Lab frame moving with constant velocity $-v\unit z$ with respect to the rest frame, the electric charge moves with velocity  $v\unit z$. In contrast to the previous sections, in this and the next sections, all quantities pertaining to the rest frame are denoted by the subscript $0$, whereas those pertaining to the Lab frame bear no such subscript. The Lorentz transformations relating the coordinates and the fields in the two frames are 
\bal
&\b r_{\bot 0} = \b r_\bot\,, & z_0= \gamma(z-vt)\,,\label{c1}\\
& E_{z0}= E_z\,, & B_{z0}=B_z\,, \label{c2}\\
& \b E_{\bot 0}= \gamma(\b E_\bot + v \unit z\b \times \b B_\bot)\,, & 
\b B_{\bot 0 }= \gamma(\b B_\bot -v \unit z\b \times \b E_\bot)\,. \label{c3} 
\gal
Here $\gamma^{-1}=\sqrt{1-v^2}$ and $\b r_\bot \cdot \unit z = 0$.   Thus, the electric field \eq{b22} in the Lab frame transforms into 
\bal
& E_{z}= E_{z0}\,, & B_{z}=0\,, \label{c5}\\
& \b E_\bot = \gamma \b E_{\bot 0}\,, & \b B_\bot = \gamma v \unit z\b \times \b E_{\bot 0}\,.\label{c6}
\gal
Using the cylindrical coordinates one can write
\ball{c8}
&\b E= E_\bot \unit b + E_z\unit z\,, & \b B= B\unit \phi\,,
\gal
where 
\bal
E_z(\b r, t)= E_0\left(\sqrt{b^2+\gamma^2(z-vt)^2},\gamma(t-vz)\right)\frac{\gamma(z-vt)}{\sqrt{b^2+\gamma^2(z-vt)^2}}\,, \label{c12}\\
B(\b r, t)= E_0\left(\sqrt{b^2+\gamma^2(z-vt)^2},\gamma(t-vz)\right)\frac{v\gamma  b}{\sqrt{b^2+\gamma^2(z-vt)^2}}\,, \label{c13}
\gal
and $E_\bot = B/v$. $E_0(r_0,t_0)$ is the magnitude of the electric field in the rest frame.  Impact parameter $b$ is a distance from the moving charge in the transverse direction.

\section{Results}\label{sec:d}

To compute the electromagnetic field created by a relativistic wave packet, one needs to specify 
the initial wave function. We adopted a Gaussian distribution   
\ball{b3}
\psi_{\b k}(0)= \frac{a^{3/2}}{\pi^{3/4}}e^{-a^2k^2/2}\,.
\gal
Its width in the coordinate space is  fixed at $a=1$~fm reflecting the strong interactions range. We also assumed that the momentum wave function is independent of $\lambda$, i.e.\ it is the same for all spin states. Other parameters used in our numerical calculations were the valence quark mass $m=0.3$~GeV, electric charge  $e$ (we omit  the quark charges 2/3 and -1/3) and the boost factor  $\gamma=100$.

\begin{figure}[ht]
\begin{tabular}{cc}
      \includegraphics[height=6cm]{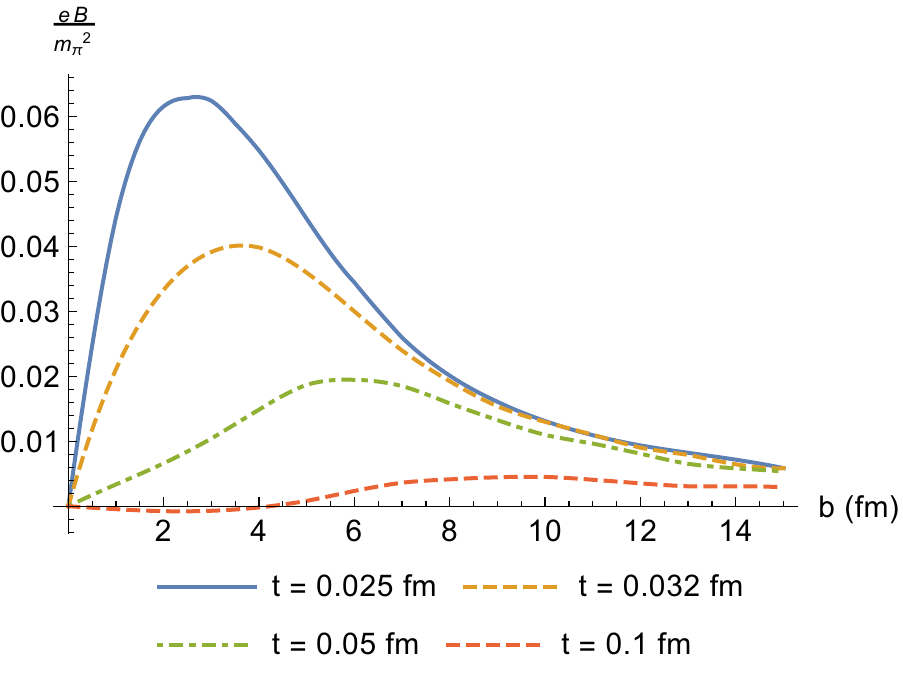} &
     \includegraphics[height=6cm]{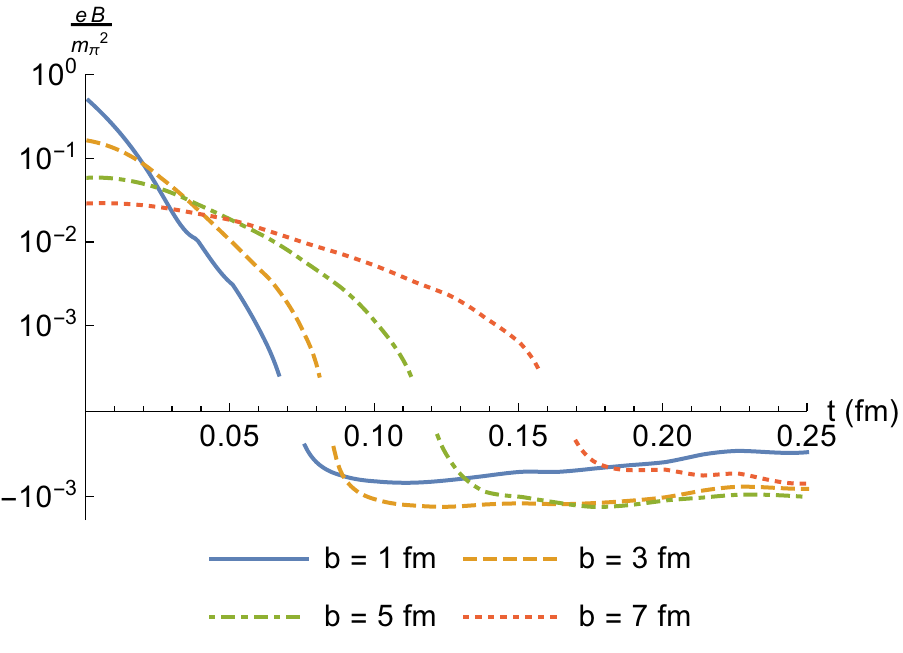}\\
       \end{tabular}
  \caption{ Magnetic field generated by a wave packet of  width $a=1$~fm as a function of impact parameter $b$ (left panel) and time $t$ (right panel). Notice, that in the right panel, there is no discontinuity of the magnetic field, as might appear at first sight. It is an artifact of the logarithmic scale on the  vertical axis, see \fig{Bvt}.}
\label{All}
\end{figure}

\begin{figure}[ht]
\begin{tabular}{cc}
      \includegraphics[height=6cm]{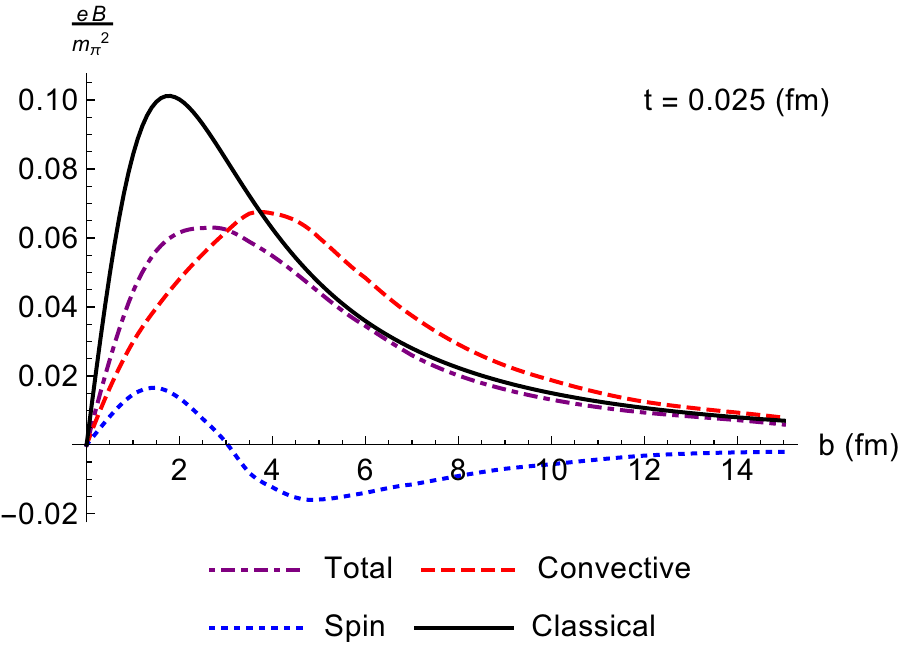} &
      \includegraphics[height=6cm]{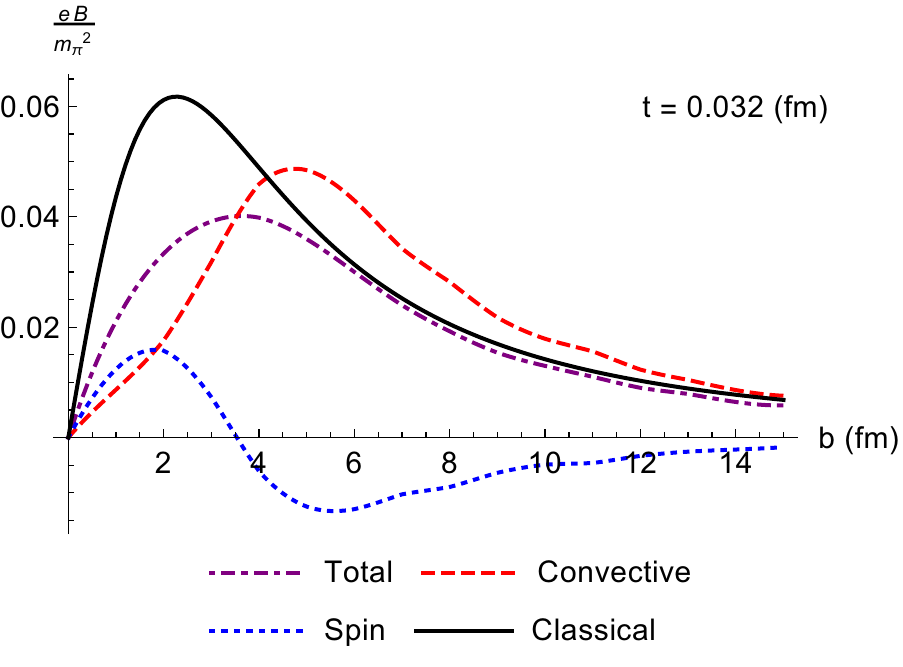}\\
      \includegraphics[height=6cm] {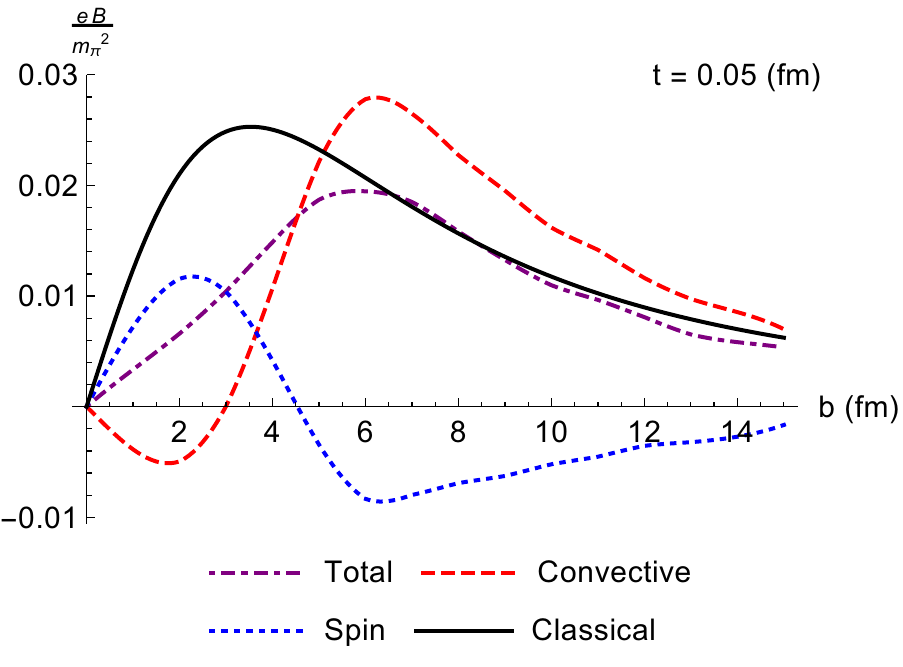}&
      \includegraphics[height=6cm]{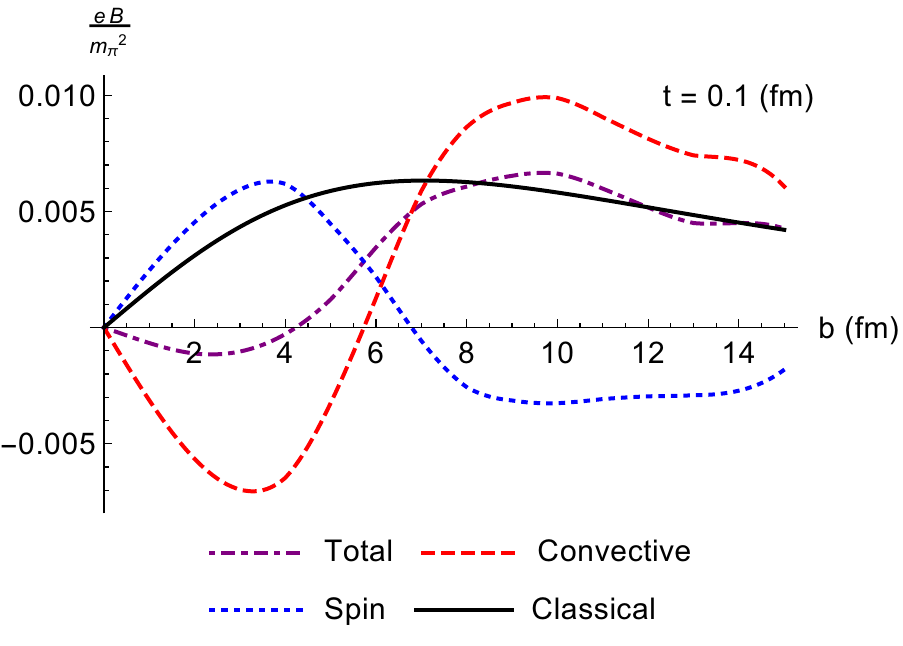}
      \end{tabular}
  \caption{Dependence of magnetic field on the transverse distance from the charge $b$ at different times.  }
\label{Bvb}
\end{figure}

\begin{figure}[ht]
\begin{tabular}{cc}
      \includegraphics[height=6cm]{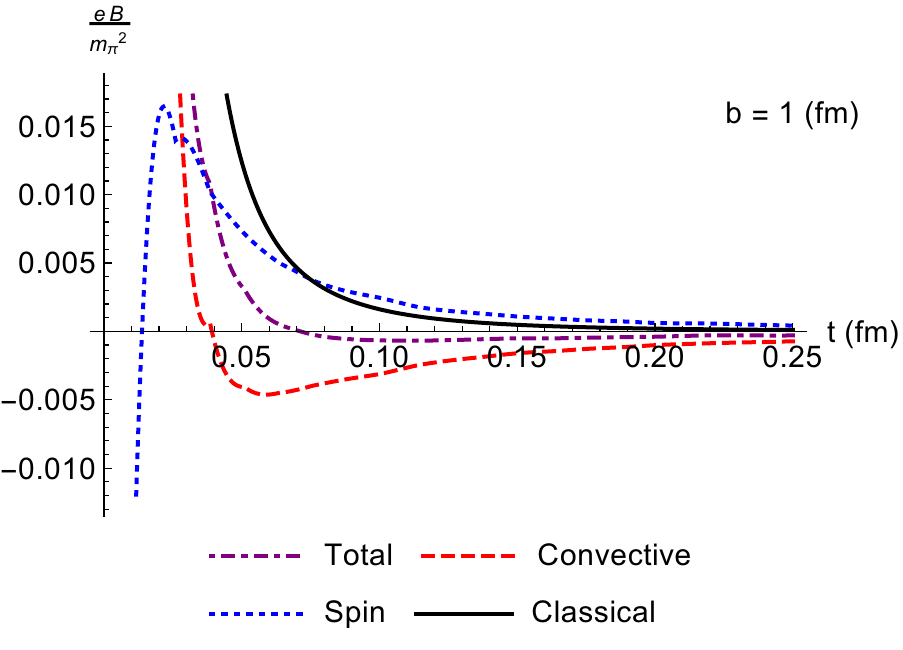} &
     \includegraphics[height=6cm]{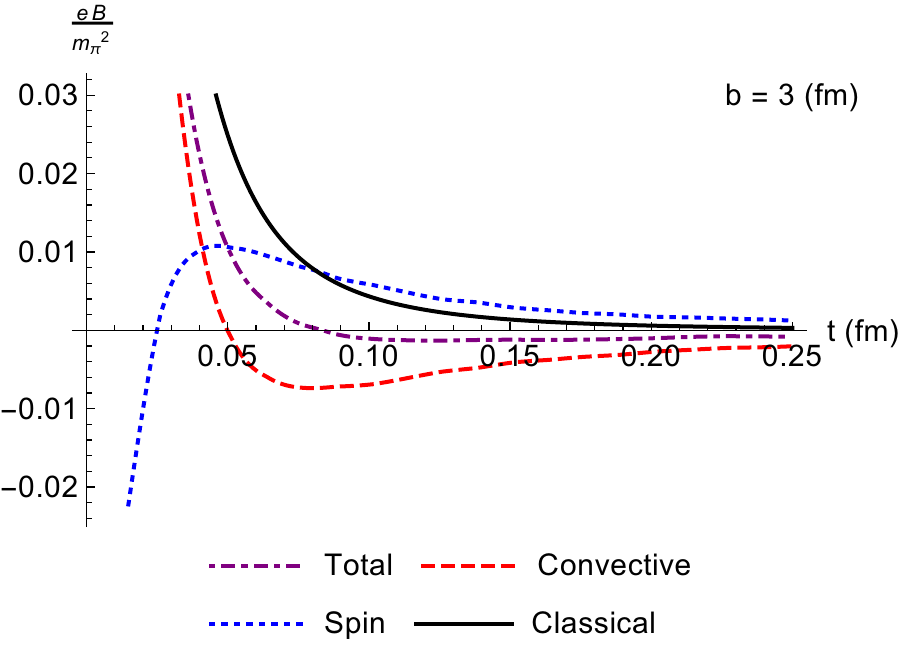}\\
     \includegraphics[height=6cm]{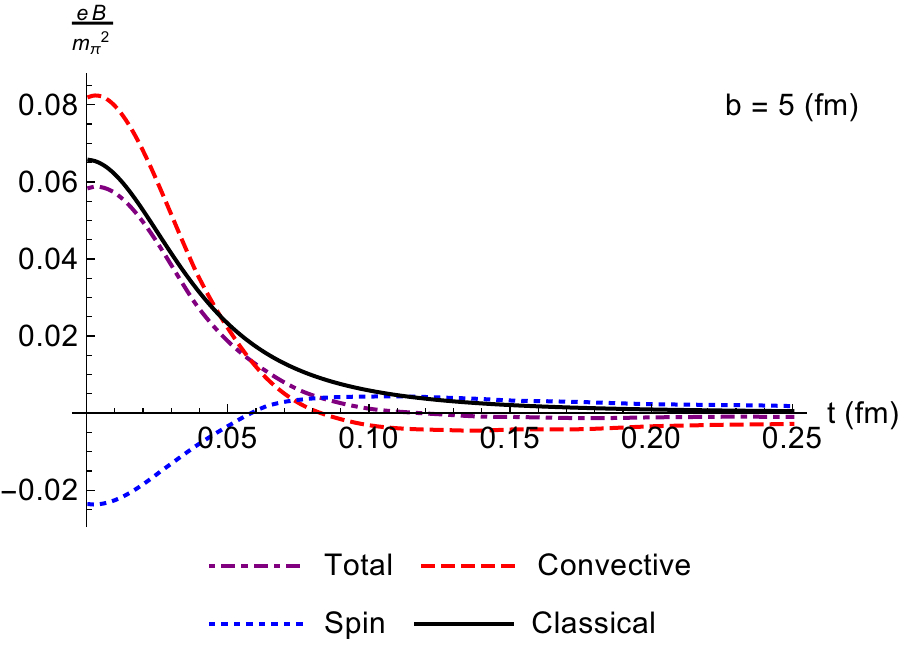} &
     \includegraphics[height=6cm]{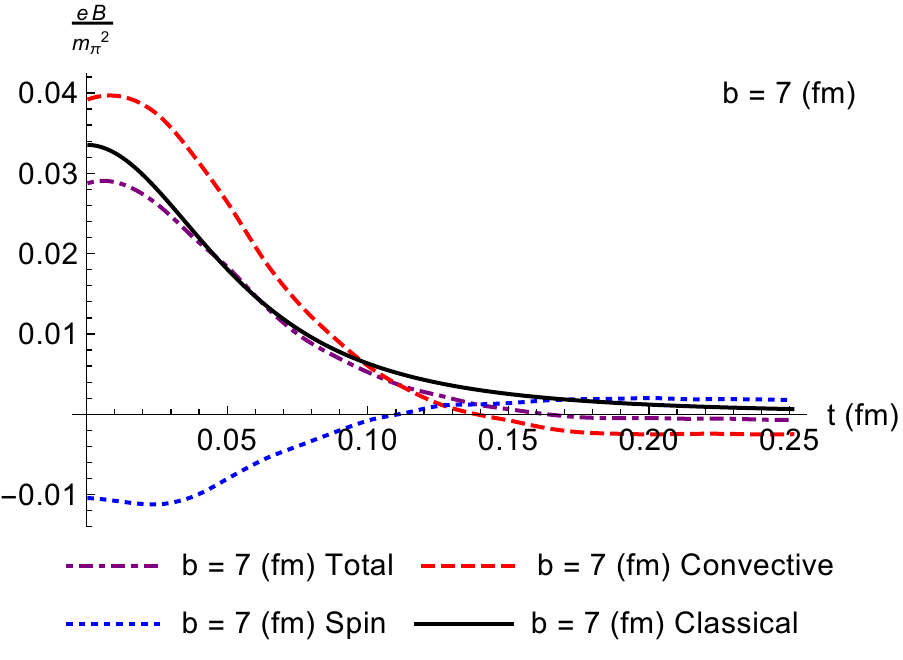}\
      \end{tabular}
  \caption{Dependence of magnetic field on time $t$ at different transverse distances $b$ from the charge.}
\label{Bvt}
\end{figure}

The results of our calculations are exhibited in \fig{All}-\fig{Bvt}.  Shown in \fig{All} is the dependence of the magnetic field on time $t$ (left panel) and distance $b$ (right panel). On the left panel notice that at later time the magnetic field changes sign. We have observed this effect before for a scalar particle \cite{Holliday:2016lbx}, where it is, in fact, much more prominent.  We will add a few more comments about the sign flip later in this section.

In \fig{Bvb} we plot each of the lines shown in \fig{All} (left panel) separately along with its convective and spin components. Also plotted is the corresponding classical (boosted Coulomb) field for comparison. At large $b$, the classical (i.e.\ point) and quantum (i.e.\ wave packet) sources induce the same field as expected. While the convective current contributes to the monopole term falling off as $1/b^2$ at large $b$, the leading spin current contribution starts with the dipole term which falls off as $1/b^3$. At later times, due to the quantum diffusion, the deviation from the classical field is observed in a wider range of distances.

In \fig{Bvt} we plot each of the lines shown in \fig{All} (right panel) separately along with its convective and spin components as well as the classical field. In order to better demonstrate the sign flip dynamics we use the linear scale for the vertical axis.  It is seen that while the convective part of the magnetic field changes its sign from positive to negative, the spin part changes its sign from negative to positive at about the same time. This makes the sign flip effect for the magnetic field generated by spin-1/2 particle less pronounced than for the field generated by a scalar particle (computed in \cite{Holliday:2016lbx}). The origin of the spin flip can be traced back to \eq{b20} in which the last two terms are proportional to the rate of decrease of charge density and the corresponding increase of the current density due to the quantum diffusion of the wave packet in the outward radial direction.

\section{Conclusions}\label{sec:e}

We computed the magnetic field created by a single valence quark, which is represented by a wave packet satisfying the Dirac equation and moving in free space with relativistic velocity. We observed that the classical  description of the valence quarks as point-particles is not accurate for calculations of the electromagnetic field in hadron and/or nuclear collisions as it breaks down at distances as large as $6$~fm at $\gamma=100$. Moreover it misses an important spin-flip effect that occurs due to the quantum diffusion of the wave packet. 

This paper is a step towards our ultimate goal of determining the structure and dynamics of electromagnetic fields created in relativistic hadron and nuclear collisions, and in particular, in heavy-ion collisions. In the forthcoming publication, we intend to study the effect of the conducting medium on the electromagnetic field created by quantum sources.

\acknowledgments
  This work  was supported in part by the U.S. Department of Energy under Grant No.\ DE-FG02-87ER40371.



\end{document}